\documentclass[12pt]{article}

\usepackage[margin=1in]{geometry}
\usepackage{setspace}
\usepackage{amsmath,amssymb,amsthm}
\usepackage{graphicx}
\usepackage{booktabs}
\usepackage[style=authoryear-comp,sorting=none,natbib=true,backend=bibtex,maxcitenames=2]{biblatex}
\usepackage[colorlinks=true, citecolor=blue, linkcolor=blue, urlcolor=blue]{hyperref}
\usepackage{subcaption}
\usepackage{microtype}
\usepackage{xcolor}

\newif\ifwithSI
\withSItrue 
\theoremstyle{plain}

\theoremstyle{remark}

\title{
 Comment on \textit{Scientific production in the era of large language models}
\\[0.5em]
\large Outcome-Triggered Treatment Timing and Spurious Event-Study Dynamics
}

\author{Thomas Renault\thanks{Université Paris-Saclay, RITM, Faculté DEM, 54 Boulevard Desgranges, 92330 Sceaux, France. Email: \textit{thomas.renault1@universite-paris-saclay.fr}} \quad Antonin Bergeaud\thanks{HEC Paris, 1 rue de la libération, 78350 Jouy-en-Josas, France. Email : bergeaud@hec.fr} \quad Clément Bosquet\thanks{Universit\'{e} Paris 1 Panth\'eon-Sorbonne, CNRS, Centre d'\'Economie de la Sorbonne, 106-112 bd de l'Hôpital, FR-75642 Paris, France. Email: clement.bosquet@univ-paris1.fr}}

\begin{document}

\maketitle

%\begin{center}
  %{\large\bfseries Outcome-Triggered Treatment Timing and Spurious Event-Study Dynamics}\\[0.3em]
  %{\large\bfseries A Comment on Kusumegi et al. (2025)}\\[1em]
  %Thomas Renault,$^{1*}$ Antonin Bergeaud,$^{2}$ Clément Bosquet$^{3\dagger}$\\[0.4em]
  %\textit{$^{1}$Université Paris-Saclay \quad $^{2}$HEC Paris \quad $^{3}$Université Paris~1 Panthéon-Sorbonne, CES}\\[0.4em]
  %$^{*}$Correspondence: \textit{thomas.renault1@universite-paris-saclay.fr}
%\end{center}

%\vspace{1em}
%\noindent\rule{\linewidth}{0.4pt}
%\vspace{0.5em}

\noindent\textbf{Abstract.} \citet{kusumegi2025} study whether researchers' preprint output rises after adopting large language models (LLMs), dating adoption as the first month in which at least one submitted abstract exceeds an LLM-detection threshold. We show that this treatment-timing rule is mechanically related to output. The probability that at least one paper is flagged in a month is increasing in the number of papers submitted in that month, so detected-adoption months are disproportionately high-output months. An event study centered on first detection can therefore display positive post-event dynamics even when the flagging rule contains no information about true LLM adoption, because the omitted pre-treatment period is selected from months with no prior detection. We demonstrate this in a simulation: with i.i.d. productivity and no causal effect, first-detection timing generates a spurious positive post-treatment path. We also replicate the stacked event study of \citet{kusumegi2025} and show that three placebo exercises (random paper-level assignment, neutral keyword flags, and a pre-ChatGPT observation window) each produce a similarly positive post-treatment pattern.

%\vspace{0.5em}
%\noindent\rule{\linewidth}{0.4pt}
%\vspace{1em}

\newpage

\section*{Introduction}

\citet{kusumegi2025} ask whether LLM adoption raises researchers' preprint output. Their design dates adoption as the first month in which at least one submitted abstract is classified as LLM-assisted by a text-based detector, and estimates the effect using a stacked difference-in-differences (DiD) event study. We show that this treatment-timing rule generates spurious event-study dynamics regardless of whether LLMs have any true effect.

We make this point two ways. First, a simulation with i.i.d. productivity and no causal effect shows that first-detection timing produces a spurious positive post-treatment path. Second, we replicate the stacked-PPML event study of \citet{kusumegi2025} from raw arXiv metadata and show that the same pattern emerges under three placebo exercises --- random paper-level flags, neutral keyword flags, and a pre-ChatGPT observation window --- each of which preserves the first-detection timing rule while removing any signal of LLM adoption.

%%─────────────────────────────────────────────────────────────────────────────
\section*{Identification problem and simulation}

\paragraph{Why the design is problematic.} Let $y_{it} \geq 0$ denote the number of papers submitted by researcher $i$ in calendar month $t$, and let $p \in (0,1)$ be the probability that any individual paper is flagged by the detector, assumed independent across papers. The probability that \emph{at least one} paper is flagged in month $t$  (i.e.\ that a detection event occurs) is
\[
q(n) \equiv \Pr(\text{detection in month }t \mid y_{it}=n) = 1-(1-p)^{n},
\]
which is strictly increasing in $n$. The treatment date used in \citet{kusumegi2025} is the \emph{first month with a detection},
\begin{equation}
  t_i^* = \min\!\bigl\{t > T_0 : \text{detection event for author }i\text{ in month }t\bigr\},
  \label{eq:first_detection}
\end{equation}
where $T_0$ denotes the introduction date of ChatGPT (December 2022).

Two consequences follow. First, because $q(n)$ is increasing in output, authors are more likely to enter treatment in months with many submissions. Treatment timing is therefore mechanically correlated with publication volume, even if the detector carries no information about true LLM adoption. Second, and more subtly, because $t_i^*$ is the \emph{first} detection event, all pre-treatment months are constrained by a ``no earlier detection'' condition: treated authors, by construction, produced no flagged paper before $t_i^*$. Post-treatment months are subject to no such constraint. The event-study comparison is therefore asymmetric in a way that mimics a causal effect: the omitted pre-treatment reference period is drawn from an unusually detection-free history, while the post-treatment period is not. A positive post-treatment path can arise from this asymmetry alone, with no causal effect and no persistence in productivity. This reflects a combination of size-biased sampling and stopping-time selection: months selected by the detection hazard are drawn from a distribution of output tilted toward high realizations, while the conditioning that defines the pre-treatment period tilts the distribution in the opposite direction. In the Supplementary Materials, we show how this bias naturally emerges in a simplified, related version of the identification.

\begin{figure}[ht]
  \centering
  \includegraphics[width=0.75\linewidth]{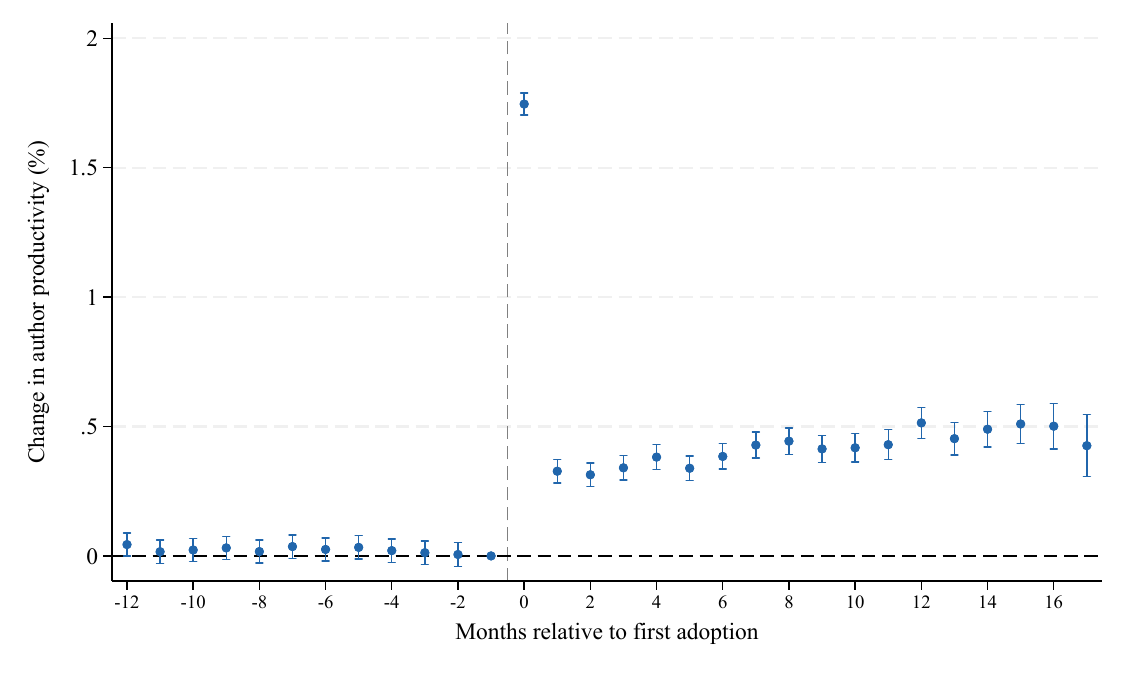}
  \caption{\textbf{Reproduction of Figure~1A from the replication package of \citet{kusumegi2025}, with the coefficient for treatment month $k=0$ plotted rather than omitted.} The visible jump at $k=0$ shows directly that treatment timing is associated with unusually high output in the treatment month itself.}
  \label{fig:kusumegi_month0}
\end{figure}

\paragraph{Treatment month in the original replication package.} The mechanical link between treatment timing and output is directly visible in the authors' replication package. Figure~1A of \citet{kusumegi2025} plots event-study coefficients around the adoption date but omits the coefficient for the treatment month itself ($k=0$). Running the same regression from the replication code and simply restoring that omitted coefficient reveals a sharp surge in output at $k=0$ (Figure~\ref{fig:kusumegi_month0}). This is precisely what the first-detection rule predicts: an author can only enter treatment in a month where they submitted at least one paper, and the more papers they submitted, the more likely that at least one is flagged. The first treatment month is therefore not a generic month but a mechanically selected high-output month. 

\begin{figure}[ht]
  \centering
  \includegraphics[width=0.85\linewidth]{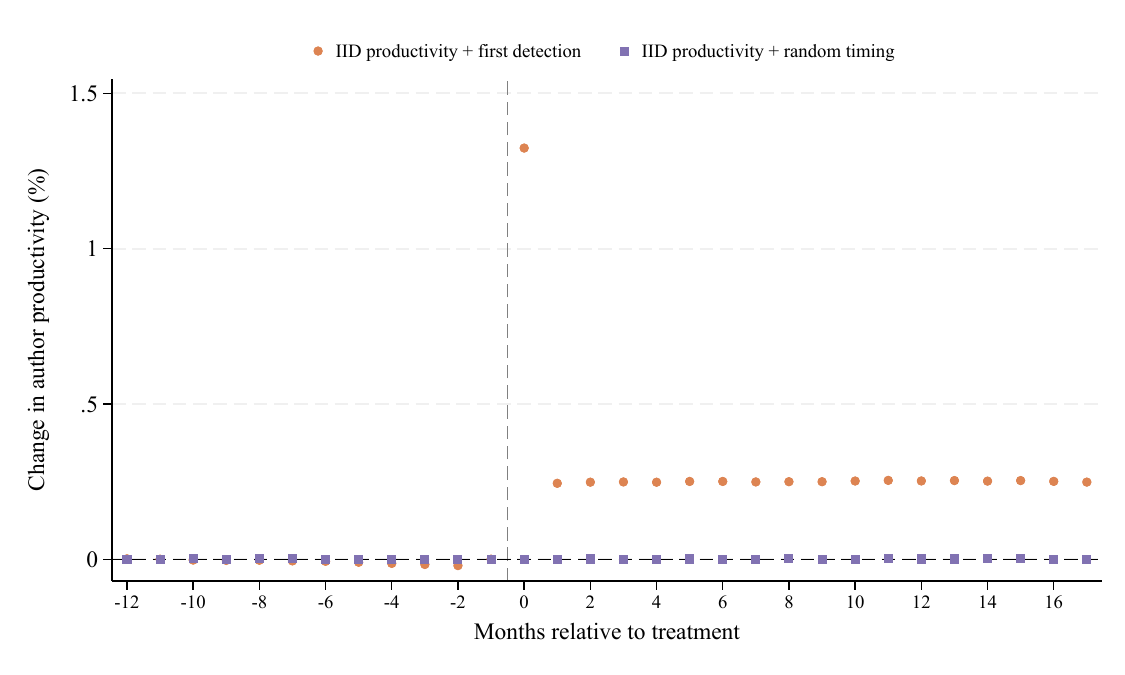}
    \caption{\textbf{Simulation of the identification problem.} Average stacked-PPML event-study coefficients across 1,000 simulations with 100,000 authors and i.i.d. productivity. Orange points reflect first-detection timing, whereas purple points are based on timing assigned randomly and independently of output. Error bars denote Monte Carlo 95\% confidence intervals for the mean coefficient at each event time, but are too small to be visible at the scale of the figure.}
  \label{fig:simulation}
\end{figure}

\paragraph{Simulation.} We illustrate this mechanism in a simulation that contains no causal treatment effect (see Materials and Methods). Productivity is independently and identically distributed across months around an author-specific mean. We compare two scenarios estimated with the same stacked-PPML specification: one in which treatment timing is defined by first detection (each manuscript independently generates a detection event with a uniform probability and an author is treated in the first month in which at least one of their manuscripts is flagged), and one in which treatment timing is assigned randomly and independently of output. Figure~\ref{fig:simulation} reports the average event-study coefficients $\gamma_k$ across 1,000 repeated simulations, where $\gamma_k$ denotes the stacked-PPML coefficient on the event-time indicator $\mathbf{1}\{t-t_i^*=k\}$, estimated with author and month-by-cohort fixed effects and $k=-1$ as the omitted reference period. 

The contrast is clear. When timing is assigned independently of output, the event-study coefficients remain centered near zero. When timing is defined by first detection, the same estimator produces positive and persistent post-treatment coefficients despite the absence of any causal effect. At the nominal 5\% level, the conventional clustered $t$-test rejects $H_0:\gamma_k=0$ in favor of a positive effect in 99.8\% of cases for $k\in[1,17]$ under first-detection timing, compared with 4.9\% under random timing. Similar results are obtained when using other modern DiD estimators for staggered designs\footnote{See for reviews \citet{dCdH_2023} and \citet{Roth_etal_2023}.} instead of stacked estimators: the issue does not stem from the estimation method itself, but from how the treatment is defined.

\section*{Replication and placebo exercises}

\begin{figure}[ht]
  \centering
  \begin{subfigure}[t]{0.48\textwidth}
    \centering
    \includegraphics[width=\linewidth]{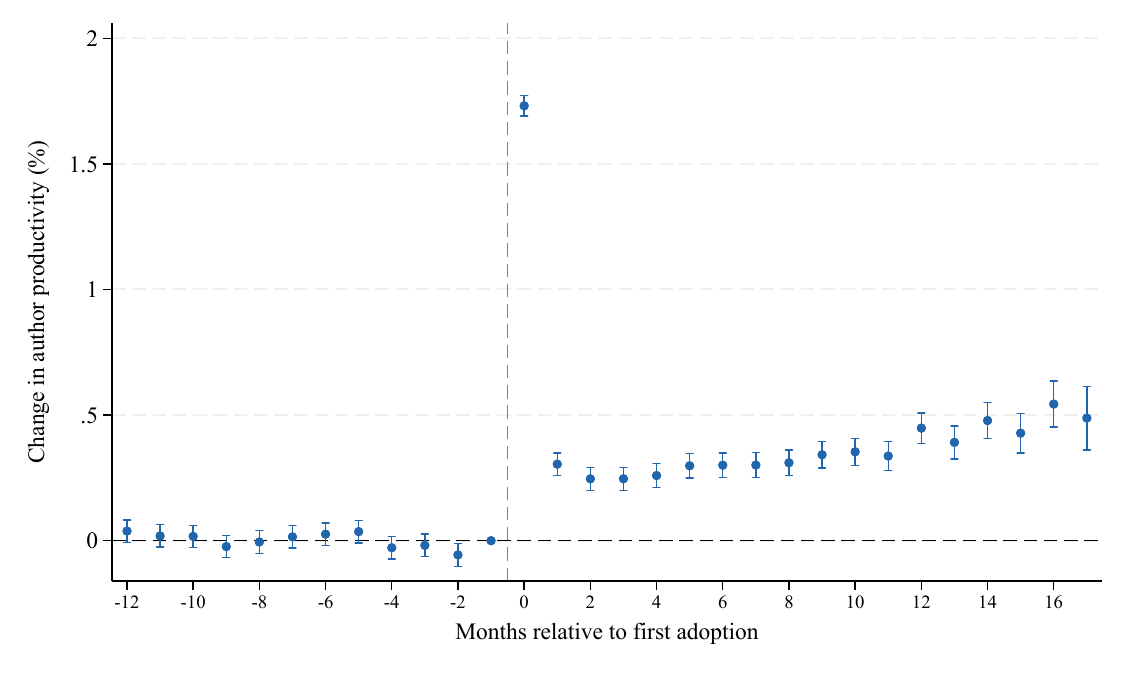}
    \caption{\textbf{Baseline replication.}}
    \label{fig:replication}
  \end{subfigure}
  \hfill
  \begin{subfigure}[t]{0.48\textwidth}
    \centering
    \includegraphics[width=\linewidth]{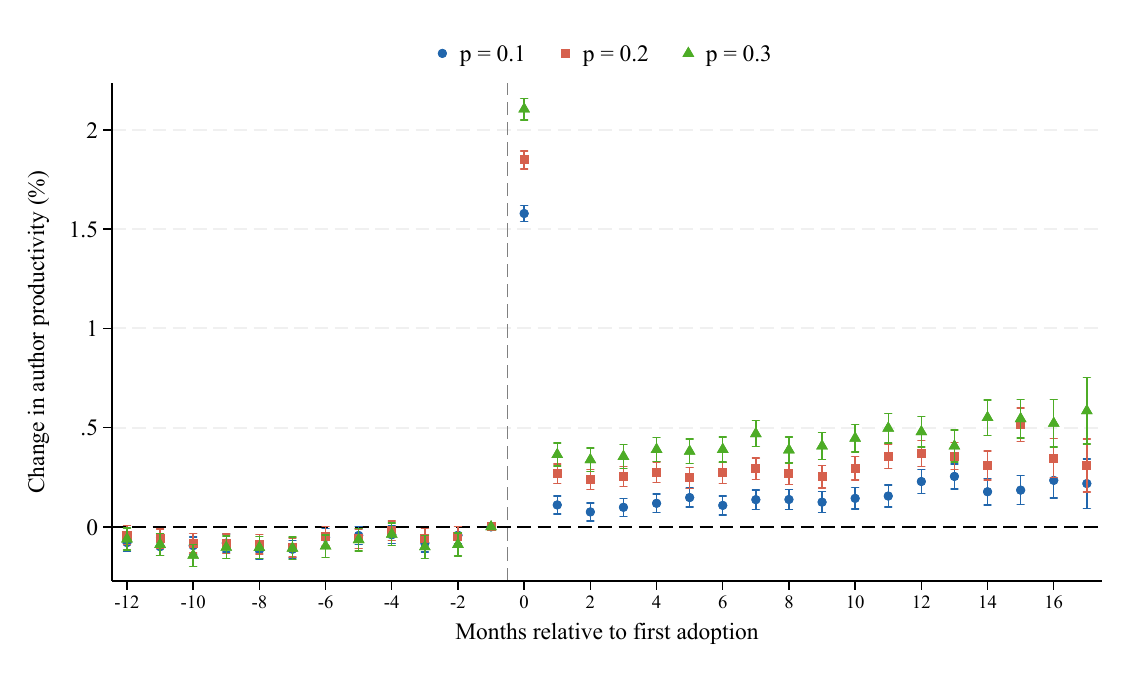}
    \caption{\textbf{Placebo 1: random assignment.}}
    \label{fig:placebo_random}
  \end{subfigure}

  \bigskip

  \begin{subfigure}[t]{0.48\textwidth}
    \centering
    \includegraphics[width=\linewidth]{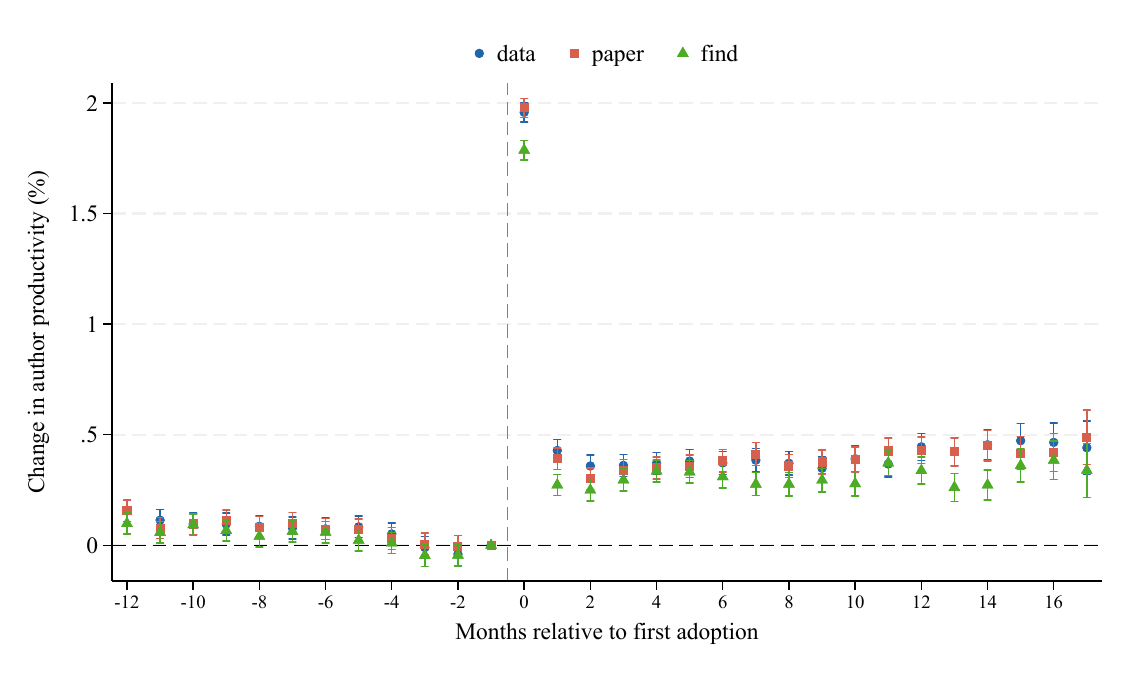}
    \caption{\textbf{Placebo 2: neutral keyword rule.}}
    \label{fig:placebo_keyword}
  \end{subfigure}
  \hfill
  \begin{subfigure}[t]{0.48\textwidth}
    \centering
    \includegraphics[width=\linewidth]{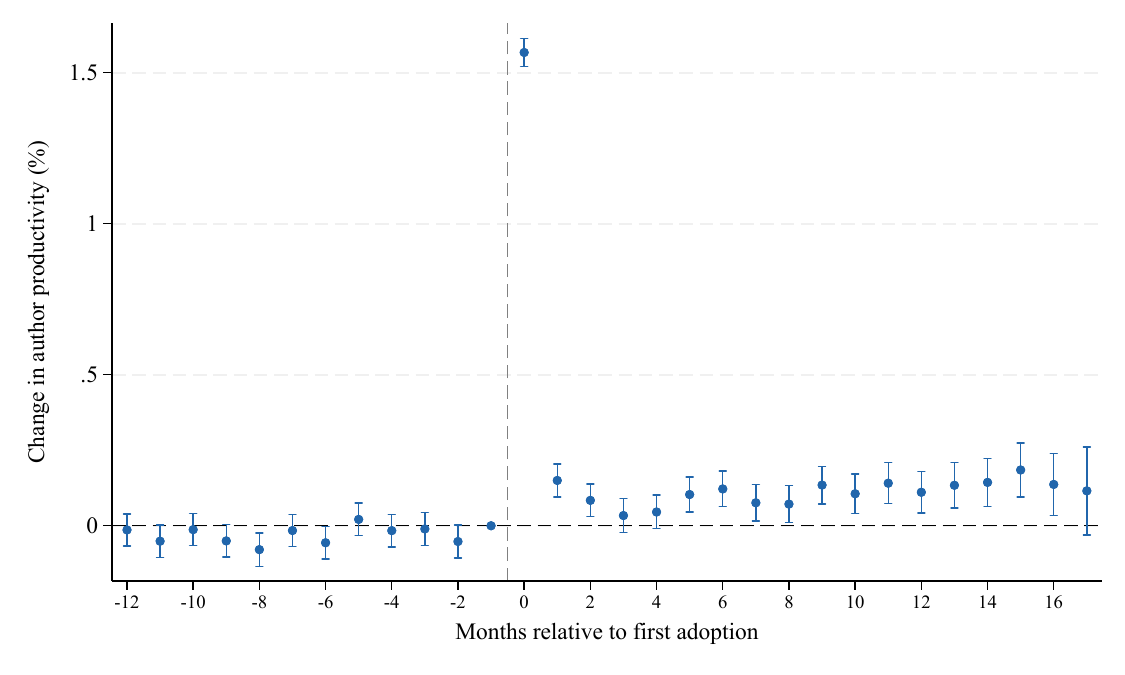}
    \caption{\textbf{Placebo 3: pre-ChatGPT window.}}
    \label{fig:placebo_prechatgpt}
  \end{subfigure}
  \caption{\textbf{Replication and placebo event studies.} All panels report stacked-PPML event-study coefficients $\hat{\gamma}_k$, with 95\% confidence intervals. Panel~A is the baseline replication; panels~B--D are placebo exercises that replace the LLM detector with random paper-level flags, neutral keyword flags, and an earlier observation window, respectively.}
  \label{fig:replication_placebo}
\end{figure}

The simulation establishes that the first-detection rule can generate spurious event-study dynamics by construction. We now ask whether the same pattern is visible in real data. We proceed in two steps. First, we reconstruct the methodology of \citet{kusumegi2025} as closely as possible: we implement their LLM detector, build the author panel from raw arXiv metadata, and estimate the same stacked-PPML event study. Our replication recovers a large positive post-treatment pattern with the same broad shape as the published result (Figure~\ref{fig:replication_placebo}A).\footnote{Small differences likely reflect variations in the detection step (the exact training sample of GPT-3.5 rewrites and the prompt used in the original study are not publicly available), as well as possibly other undocumented preprocessing choices. The replication package of \citet{kusumegi2025} only includes the aggregated panel (at the year-author level, with hashed anonymized authors), but not the paper-level data.} Second, we apply three placebo procedures. Each replaces the LLM detector with a flagging rule that carries no information about true LLM adoption, while keeping the first-detection timing structure identical.

\paragraph{1. Random paper-level assignment.} Each paper posted after December 2022 is independently flagged with probability $p$ — drawn uniformly and without reference to abstract content — and an author's treatment date is defined as the first month in which at least one of their papers receives a flag. Since flagging is uninformative by construction, any post-treatment pattern must reflect the timing rule alone (Figure~\ref{fig:replication_placebo}B).

\paragraph{2. Neutral keyword placebo.} A paper is flagged if its abstract contains the neutral keywords \textit{data}, \textit{paper}, or \textit{find}. An author's placebo event time is defined as the first month in which at least one such paper appears after December 2022. (Figure~\ref{fig:replication_placebo}C).

\paragraph{3. Pre-ChatGPT placebo window.} We shift the entire analysis to a period before ChatGPT existed: the observation window runs from January 2020 to June 2022. The same LLM detector and stacked-PPML specification are applied unchanged. Any positive post-treatment pattern here cannot be attributed to LLM adoption, since the technology was not yet available (Figure~\ref{fig:replication_placebo}D).

All three placebos recover a positive post-treatment event-study path of similar shape and magnitude to the baseline replication in Panel~A. The average post-treatment coefficient ($\bar\gamma_k$ for $k\in[1,17]$) is 0.36 in the baseline replication, 0.37 in the keyword placebo\footnote{0.40, 0.39 and 0.31 for the words \emph{data}, \emph{paper} and \emph{find}, respectively.}, 0.30 in the random-flag placebo\footnote{0.16, 0.31 and 0.44 with $p=0.1, 0.2 \text{ and } 0.3$, respectively.}, and 0.11 in the pre-ChatGPT placebo. The positive pattern is therefore not specific to the LLM detector: it emerges whenever the first-detection timing rule is applied, regardless of what is being detected. Lower detection hazard leads to smaller post-treatment effects (Figure~\ref{fig:replication_placebo}B), which also explains why the pre-ChatGPT placebo produces weaker estimates, as the probability that a paper is flagged is lower in that setting. More generally, the relationship between detection hazard and estimated effect size also explains the result in \citet[Figure S13]{kusumegi2025}, where the estimated effect declines as the mixture cutoff $\alpha$ used to classify LLM papers increases.

\section*{Discussion}

We do not argue that LLMs cannot affect scientific productivity, nor do we attempt to recover the exact decomposition of the estimate in \citet{kusumegi2025} into causal and non-causal components. Our claim is that defining treatment timing as the first month in which at least one paper is flagged mechanically links the treatment variable to publication volume. As a result, positive post-treatment event-study dynamics can arise even in the absence of any causal effect, and therefore do not constitute distinctive evidence of a productivity effect.

This concern is related to, but distinct from, the usual endogeneity of technology adoption. The authors of the original paper already note that adoption may be non-random and that timing may be endogenous to productivity. Our point is that even if one brackets those broader concerns, the act of dating treatment from observed output creates an additional design problem. The event study can appear persuasive because months selected by the detector are not generic months; they are months in which the opportunity for detection is greatest, while the omitted pre-treatment reference period is conditioned on the absence of any prior detection. This asymmetry alone is sufficient to generate the observed pattern, as our i.i.d. simulation demonstrates.

\vspace{1em}
\noindent\textbf{Acknowledgements.} We thank Gabriela Terra for research assistance.

\vspace{0.5em}
\noindent\textbf{Data and code availability.} Replication code is available in the project repository: \url{https://github.com/trenault/llm_productivity_comment}. Parts of the code were developed with assistance from Claude Code and subsequently reviewed and validated by the authors. The arXiv dataset is publicly available on Kaggle: \url{https://www.kaggle.com/datasets/Cornell-University/arxiv}.

\printbibliography

\section*{Materials and Methods}

\paragraph{Simulation design.} Our simulation generates a 30-month panel of author-level publication counts for 100,000 authors, consistent with the number of distinct authors in the true arXiv panel dataset. Monthly output for each author is drawn independently from a Poisson distribution with an author-specific mean, so productivity is i.i.d. over time by construction. There is no causal treatment effect in the data-generating process. From month 11 onward, each simulated manuscript independently receives a flag with probability p, so an author's monthly detection probability is $1-(1-p)^{y_{it}}$ where $y_{it}$ is their output that month. We compare two scenarios using the same stacked-PPML specification: first-detection timing (an author is treated in the first month with at least one flagged manuscript) and random timing assigned independently of output. The results are summarized by the average coefficient path and Monte Carlo 95\% confidence intervals across 1,000 repeated simulations.

\paragraph{Panel construction.} We retrieve arXiv metadata from the Cornell/arXiv snapshot on Kaggle, covering January 2016 to June 2024. We exclude papers in core AI subfields (cs.CV, cs.LG, cs.AI, cs.IR, cs.CL) following  \citet[SM~S1.1]{kusumegi2025}. The active-author sample for the main analysis requires at least four publications between 2018 and 2021; the observation window is January 2022 to June 2024 (30 months). For the pre-ChatGPT placebo, the activity filter requires at least four publications between 2016 and 2019, and the observation window is January 2020 to June 2022 (30 months).

\paragraph{LLM detection model.} We implement a mixture-style detector following \citet[SM~S2.3]{kusumegi2025}, but trained on a novel set of 20,000 synthetic paired abstracts. We compute unigram frequencies separately in the human abstracts and in their GPT-3.5 rewrites, retain words observed in both corpora, and form word-level log-likelihood ratios. For each abstract, sentence-level log-likelihood ratios are summed within sentence and the implied mixture likelihood is numerically maximized over $\alpha \in [0,1]$. We classify a paper as LLM-related when the estimated mixture share exceeds $0.1$. In the main replication, this classification is applied only to papers posted after December 2022. In the pre-ChatGPT placebo, the same detector is applied after the placebo date of December 2020.

\paragraph{Placebo tests.} All three placebo tests use the same stacked DiD specification as the main analysis; only the treatment indicator and, where applicable, the observation window differ.

\textit{Placebo~1 (random assignment).} Each paper published after December 2022 receives an independent $\mathrm{Bernoulli}(p)$ flag, drawn without reference to abstract content. A researcher's placebo first-treatment month is the first month in which at least one of their papers carries this random flag.

\textit{Placebo~2 (neutral keywords).} A paper is flagged if its abstract, posted after December 2022, contains \textit{data}, \textit{paper} or  \textit{find}. A researcher's placebo first-treatment month is defined as in the main analysis, with this keyword indicator replacing the LLM detection score.

\textit{Placebo~3 (pre-ChatGPT period).} We apply the same implemented detector to abstracts in a January 2020 to June 2022 observation window, using December 2020 as a placebo ChatGPT release date (instead of December 2022). The active-author filter is shifted backward as well, requiring at least four publications between 2016 and 2019. 

\paragraph{Stacked DiD.} We follow the stacked DiD design of \citet{kusumegi2025}. Each cohort is defined by the calendar month of first detection under the relevant rule. Never-treated authors are assigned pseudo-treatment months drawn uniformly from the post-introduction window: January 2023 to June 2024 in the main, random, and keyword analyses, and January 2021 to June 2022 in the pre-ChatGPT placebo. For each cohort, we stack the treated authors and the never-treated authors assigned to that cohort. We estimate a Poisson pseudo-maximum-likelihood (PPML) event-study regression with author and calendar-month$\times$cohort fixed effects, and standard errors clustered at the researcher level \citep[SM~S3.1]{kusumegi2025}. The reference period is $k=-1$. Similar results are obtained when using other modern estimators for staggered designs instead of stacked estimators.\footnote{See for reviews \citet{dCdH_2023} and \citet{Roth_etal_2023}.}

\ifwithSI
\section*{Supplementary Materials}

\subsection*{A Closed-Form Expression for the Mechanical Bias in Event Studies}

The simulation in our comment shows that first-detection timing produces a positive event-study path even when productivity is i.i.d.\ and the detector is uninformative. The mechanism admits a closed-form description as a direct consequence of size-biased sampling. For tractability, the derivation below considers a single cohort of authors who share the same treatment date, estimated by PPML with author fixed effects. This is the natural building block of the stacked estimator used in the main text: within each cohort-specific stack, the setting is exactly the one analyzed here, and month$\times$cohort fixed effects ensure that the single-cohort closed form applies.

\subsubsection*{Setup}

Let $y_{it}$ denote author $i$'s number of submissions in month $t$, drawn i.i.d.\ over time from a distribution $F$ with mean $\mu=\mathbb{E}[y]$, and let $n$ index a generic non-negative integer realization of $y$. Each submitted paper is independently flagged with probability $p\in(0,1)$. A \emph{detection event} in month $t$ is the event that at least one paper that month is flagged. Conditional on $y_{it}=n$, the detection probability is
\[
q(n) \;=\; 1-(1-p)^{n},
\]
a function that is zero at $n=0$ and strictly increasing in $n$. Treatment is dated at the first month with a detection,
\[
t_i^{*} \;=\; \min\{\,t>T_0 : \text{detection at }t\,\}.
\]

\subsubsection*{Theoretical values under the null hypothesis of no effect}

We assume no causal effect: $y_{it}$ is i.i.d.\ across months. Throughout what follows, $k_i=t-t_i^*$ denotes the signed number of months relative to author $i$'s own treatment date.  $t_i^*$ is a random stopping time whose hazard $q(y_{it})$ depends on the contemporaneous outcome. This induces three distinct conditional distributions of $y$, one for each region of event time.

\begin{itemize}
    \item \textbf{$k=0$ (treatment month)} Because the month $t_i^*$ is selected precisely because a detection occurred, and because $q(y)$ is increasing in $y$, the conditional distribution of output in the treatment month is the $q$-weighted (size-biased) distribution of $F$:
\[
\Pr\bigl(y_{i,t_i^*}=n\bigr) \;=\; \frac{q(n)\,\Pr(y=n)}{\mathbb{E}[q(y)]}.
\]
Taking expectations,
\[
\mathbb{E}\bigl[y_{i,t_i^*}\bigr]
\;=\; \frac{\mathbb{E}[y\,q(y)]}{\mathbb{E}[q(y)]}
\;=\; \mu \;+\; \frac{\mathrm{Cov}(y,q(y))}{\mathbb{E}[q(y)]}
\;>\;\mu,
\]
with strict inequality whenever $F$ has positive variance and $p>0$. Months in which a detection is observed are, on average, months with above-average output.

\item \textbf{$k=-1$ (reference period)} The month just before the first detection, by construction, contains no flagged paper. Its conditional distribution is weighted by $1-q(y)=(1-p)^y$, which is decreasing in $y$. Therefore
\[
\mathbb{E}\bigl[y_{i,t_i^*-1}\bigr]
\;=\; \frac{\mathbb{E}[y\,(1-p)^y]}{\mathbb{E}[(1-p)^y]}
\;=\; \mu \;-\; \frac{\mathrm{Cov}(y,q(y))}{1-\mathbb{E}[q(y)]}
\;<\;\mu.
\]
The reference period is selected from months in which no paper was flagged. Since the probability of at least one flag grows with the number of submissions, unflagged months are systematically low-output months.

\item \textbf{$k \ge 1$ (post-treatment periods)} Conditional on $t_i^*$, future months carry no selection: $y_{i,t_i^*+k}$ for $k_i\ge 1$ is still i.i.d.\ $F$, so $\mathbb{E}[y_{i,t_i^*+k}]=\mu$.
\end{itemize}

With author fixed effects and $k_i=-1$ as the reference period, the PPML event-study estimator identifies the log-ratio of conditional means. Combining the three expressions above yields a closed-form expression for the probability limits of the PPML event-study coefficients (under the null hypothesis of no effect):
\[
\gamma_k^{\text{\emph{null}}} \;=\;
\begin{cases}
\displaystyle \log\!\left(\dfrac{\mathbb{E}[y\,q(y)]}{\mathbb{E}[q(y)]}\Big/\dfrac{\mathbb{E}[y\,(1-p)^y]}{\mathbb{E}[(1-p)^y]}\right) & k=0,\\[1.2em]
\displaystyle \log\!\left(\mu\Big/\dfrac{\mathbb{E}[y\,(1-p)^y]}{\mathbb{E}[(1-p)^y]}\right) & k\ge 1.
\end{cases}
\]
Both expressions are strictly positive for any $p\in(0,1)$ and any non-degenerate $F$.

\subsubsection*{Working example}

When $y\sim\text{Poisson}(\mu)$, the probability-generating function $\mathbb{E}[s^y]=e^{\mu(s-1)}$ gives closed forms. Using $\mathbb{E}[(1-p)^y]=e^{-\mu p}$ and $\mathbb{E}[y(1-p)^y]=(1-p)\mu e^{-\mu p}$,
\[
\mathbb{E}[y_{i,t_i^*-1}] = \mu(1-p), \qquad
\mathbb{E}[y_{i,t_i^*}] = \mu\,\frac{1-(1-p)e^{-\mu p}}{1-e^{-\mu p}},
\]
so the expected event-study path reduces to
\[
\gamma_k^{\text{null}} \;=\; \begin{cases}
\displaystyle \log\!\left(\dfrac{1-(1-p)e^{-\mu p}}{(1-p)\,(1-e^{-\mu p})}\right) & k=0,\\[0.9em]
\displaystyle -\log(1-p) & k\ge 1.
\end{cases}
\]

We can make three comments. First, the post-treatment plateau $-\log(1-p)$ depends only on $p$ and not on the productivity distribution. Second, the bias at $k=0$ is largest for low-productivity authors: as $p\to 0$, $\gamma_0\to\log(1+1/\mu)$. Third, the full path has a characteristic shape:  a large jump at $k=0$ followed by a lower plateau at $-\log(1-p)$, precisely the shape visible in Figure~1A of \citet{kusumegi2025} once $k=0$ is restored.\footnote{The parameter $p$ is the marginal probability that a single paper triggers a detection. It is a weighted average of the detector's sensitivity and its false-positive rate, not the adoption rate itself. The mechanical bias does not vanish if one grants the authors a perfect detector: if $p$ coincides with the true adoption rate, $-\log(1-p_{\text{adoption}})$ is still the spurious component, because the bias is driven by the first-hitting-time timing rule, not by detector error. The bias vanishes only if treatment is defined without reference to a first-detection hazard on an output-dependent outcome.}
\fi

\end{document}